\documentclass[preprint,aps,showpacs,nofootinbib]{revtex4}

\def\bea{\begin{eqnarray}}
\def\eea{\end{eqnarray}}
\def\bec{\begin{center}}
\def\ec{\end{center}}
\def\beq{\begin{equation}}
\def\eeq{\end{equation}}

\begin{document}

\draft
\tighten
\preprint{KIAS-P07086}
\title{\large \bf Sequestered uplifting and the pattern of soft
\\
supersymmetry breaking terms}
\author{Kwang Sik Jeong\footnote{ksjeong@kias.re.kr}}
\address{School of Physics, Korea Institute for Advanced Study (KIAS),
Seoul 130-722, Korea}

\begin{abstract}
We examine the pattern of soft supersymmetry breaking terms in moduli stabilization,
where an uplifting potential is provided by spontaneously broken supersymmetry in
a generic sequestered sector.
From stationary conditions, we derive the relation between moduli F-term vacuum
expectation values which does not depend on the details of sequestered uplifting.
This moduli F-term relation is crucial for identifying the dominant source of soft
terms of visible fields.
\end{abstract}

\pacs{}
\maketitle

\section{Introduction}
In string compactifications, the shape and size of the internal space are
parameterized by moduli.
Since the effective couplings of low energy theory are generically determined by
moduli vacuum expectation values (VEVs), it is important to understand how
the moduli are stabilized.
It is naturally considered to add non-perturbative effects as a source of moduli
fixing, and then the moduli can be fixed well by imposing supersymmetric stationary
conditions.
The corresponding vacuum energy is, however, negative for non-vanishing gravitino
mass.
In this case, a de-Sitter (dS) vacuum can be constructed via the uplifting
mechanism \cite{Kachru:2003aw,Lebedev:2006qq,Burgess:2003ic,Choi:2006bh},
where a supersymmetry (SUSY) breaking sector is introduced to provide positive
energy to the scalar potential.

The SUSY breaking in uplifting sector would give rise to soft SUSY breaking terms,
through effective cross-couplings between visible and uplifting sector fields,
in the low energy lagrangian.
Soft terms can involve flavor or CP violations which are already restricted by
experiments.
Hence, in order to avoid an additional source of flavor violations,
the cross-couplings should be strongly suppressed unless the mediating interactions
are flavor-blind.
The suppression of such troublesome cross-couplings has led to the idea of
sequestering \cite{Randall:1998uk}.
It has been noticed that the sequestering is realized by spatially separated
branes in strongly warped compactification \cite{Choi:2006bh,Luty:2000ec}.
This geometrical sequestering can be also understood as the dual description
of conformal sequestering \cite{Schmaltz:2006qs} according to the AdS/CFT
correspondence \cite{Maldacena:1997re}.
If sequestered from the visible sector, the uplifting sector would generate a dS
vacuum without inducing additional soft terms of visible fields.
This feature is phenomenologically desirable as the addition of uplifting effects
causes no flavor violations.

In this paper, we wish to discuss the pattern of soft terms of visible fields within
the framework of 4D supergravity, where an uplifting sector is sequestered from
visible sector and spontaneous SUSY breaking takes place there.
Particularly, we focus on how to identify the dominant source of visible soft terms.
For this end, in the next section, we examine the general features of supergravity
with a sequestered sector and present the relation between SUSY breaking quantities
derived from stationary condition.
It turns out that the dominant source of visible soft terms can be determined
irrespectively of the detailed form of the sequestered sector.
In section 3, using this property, it is found that moduli stabilization by
non-perturbative effects naturally leads to mirage mediation
\cite{Choi:2004sx,Choi:2005ge,Choi:2005uz,Endo:2005uy},
while a dS vacuum is achieved by the uplifting mechanism.
Section 4 is the conclusion.

\section{Sequestered uplifting in supergravity}
In the flat spacetime background, the effective action for N=1 supergravity
coupled to the gauge and matter superfields can be written in the rigid superspace
\cite{Cremmer:1982en}
\bea
\label{action}
{\cal S} =
\int d^4x \sqrt{-g}\left[
\int d^2\theta d^2\bar\theta C\bar C (-3e^{-K/3})
+ \left\{ \int d^2\theta
\left(\frac{1}{4}f_aW^{a\alpha}W^a_\alpha + C^3W\right)
+ {\rm h.c.} \right\} \right],
\eea
where $K$, $W$ and $f_a$ denote the K\"ahler potential, superpotential and
gauge kinetic functions, while $C=C_0+\theta^2F^C$ is the non-physical chiral
compensator for super-Weyl invariance.
The SUSY breaking effects originated from gravity are represented by the
chiral compensator which involves the scalar auxiliary field of supergravity
multiplet.
Meanwhile, in the presence of radiative corrections, the super-Weyl invariance
is maintained as a result of non-trivial dependence on $C$ of running couplings.
This implies that the F-term of chiral compensator generates soft terms at
loop level \cite{Randall:1998uk,Giudice:1998xp,Pomarol:1999ie}.
Since soft terms always receive such contributions from anomaly
mediation, it is quite important to know the relative importance of auxiliary
components of SUSY breaking fields compared to $F^C$.
In the Einstein frame where the graviton kinetic term is canonical,
SUSY breaking quantities \cite{Pomarol:1999ie} are given by
\bea
\label{FD-terms}
F^I &=& -e^{K/2}K^{I\bar J}(D_J W)^*, \qquad
\frac{F^C}{C_0} \,=\, \frac{1}{3}(\partial_I K) F^I + e^{K/2}W^*,
\nonumber \\
D^a &=& -\frac{1}{{\rm Re}(f_a)} \eta^I_a\partial_I K,
\eea
where $D_IW=\partial_I W+(\partial_I K)W$ is the K\"ahler covariant derivative
of superpotential, and $\eta^I_a$ denote the holomorphic Killing vectors for
the infinitesimal gauge transformation of $\Phi^I$,
i.e. $\delta \Phi^I = \Lambda^a \eta^I_a(\Phi)$ for holomorphic $\Lambda^a$.
Their VEVs are determined by minimizing the supergravity scalar potential
$V_{\rm tot}=V_F+V_D$,
\bea
V_F &=& e^K \left\{ K^{I\bar J}(D_IW)(D_JW)^* -3|W|^2 \right\}
= K_{I\bar J}F^IF^{J*} - 3|m_{3/2}|^2,
\nonumber \\
V_D &=& \frac{1}{2}{\rm Re}(f_a) D^aD^a,
\eea
in which $m_{3/2}=e^{K/2}W$ is the gravitino mass.
In supergravity, supersymmetric field configurations satisfying $D_IW=0$
correspond to a stationary point of $V_{\rm tot}$\footnote{
Due to the gauge invariance, the D-terms can be written in terms of K\"ahler
covariant derivatives as $D^a=-\frac{1}{{\rm Re}(f_a)}\frac{\eta^I_a D_IW}{W}$.
Thus, it is obvious that the field configuration satisfying $D_IW=0$ leads to
$F^I=D^a=0$ for $W\neq0$ \cite{Choi:2005ge}.}
if the superpotential does not vanish.
However, the supersymmetric vacua have a negative vacuum energy, which should be
compensated by SUSY breaking effects in order to achieve a phenomenologically
viable dS or Minkowski vacuum.

In the presence of a sequestered sector, the effective supergravity is
generically specified by K\"ahler and superpotential taking the form
\bea
\label{sequestering}
K &=& -3\ln\Omega = -3\ln\left( \Omega_{\rm vis}(Q,\bar Q,T^i,\bar T^i)
+ \Omega_0(T^i,\bar T^i)
+ \Omega_{\rm seq}(Z^\alpha,\bar Z^\alpha,\tilde V^a) \right),
\nonumber \\
W &=& W_{\rm vis}(Q,T^i) + W_0(T^i) + W_{\rm seq}(Z^\alpha),
\eea
where $Q$ stands for the matter superfields in visible sector, and $T^i$ denote
the moduli whose VEVs determine the effective couplings of visible fields,
while $Z^\alpha$ and $\tilde V^a$ are matter and vector superfields living
in the sequestered sector respectively.
In the superconformal frame, soft terms of visible fields are generated through
effective cross-couplings between the visible and SUSY breaking fields
\cite{Randall:1998uk}.
From the above sequestered structure, however, it is manifest that the fields
in sequestered sector do not have cross-couplings with other sector fields
in the superconformal frame lagrangian (\ref{action}).
Hence the auxiliary components of $Z^\alpha$ and $\tilde V^a$ are irrelevant
to visible soft terms, but have uplifting effects which would be necessary
to realize a dS vacuum.
Despite the absence of cross-couplings with other sector fields, the sequestered
sector still influences the moduli configuration minimizing supergravity scalar
potential through gravitational effects.
This can be understood from that the chiral compensator couples to any operator
in K\"ahler and superpotential.
Indeed, the equation of motion for moduli $T^i$ is generally not decoupled from
the sequestered sector if supersymmetry is broken, though the supersymmetric
stationary condition corresponds simply to $D_iW=D_\alpha W=0$.
For the study of SUSY breaking, one needs to know the VEV of SUSY breaking
quantities rather than the field configuration itself at a vacuum.
In this view, it is worthwhile to rephrase the stationary conditions in terms
of auxiliary components
\bea
\partial_I V_F &=&
-2\frac{\partial_I \Omega}{\Omega}V_F
- \left( \frac{\partial_I\partial_J W}{W} F^J
+ 2\frac{\partial_I W}{W} \frac{F^C}{C_0}  \right)m_{3/2}
\nonumber \\
&& +\,
3\frac{\partial_I\partial_J\partial_{\bar L}\Omega}{\Omega}F^JF^{L*}
+ 3\frac{\partial_I\partial_J \Omega}{\Omega}
F^J\left(\frac{F^C}{C_0}\right)^*,
\nonumber \\
\partial_I V_D &=&
-\left(\partial_I \ln({\rm Re}(\tilde f_a))
+ 2\frac{\partial_I \Omega}{\Omega} \right)V_D
+ 3 \frac{\tilde \eta^{J *}_a\partial_I\partial_{\bar J}\Omega}{\Omega}
\tilde D^a,
\eea
where $I=\{i,\alpha\}$, $V_D$ is the D-term scalar potential for $\tilde V^a$,
and we have chosen the Einstein frame condition.
For the gauge interactions in sequestered sector, the associated gauge couplings
have no dependence on moduli $T^i$ and only the matter superfields living there
are charged under the gauge group:
\bea
\tilde f_a = (\mbox{$T^i$-independent function}), \qquad
\tilde \eta^i_a = 0, \qquad
\tilde \eta^\alpha_a = \tilde \eta^\alpha_a(Z^\beta),
\eea
where $\tilde f_a$ and $\tilde\eta^I_a$ are the gauge kinetic function and
holomorphic Killing vectors of sequestered sector gauge group, respectively.
The sequestering therefore leads to that, written in terms of auxiliary components,
the stationary conditions have the decoupled structure.
Concretely, from $\partial_i V_{\rm tot}=\partial_i(V_F+V_D)=0$, the sequestered
structure (\ref{sequestering}) results in the relation between moduli F-term VEVs
\bea
\label{Moduli-F-term-relation}
\frac{1}{3}\frac{\partial_i\partial_j W_0}{W}
\hat F^j
+ \frac{2}{3}\frac{\partial_i W_0}{W}\hat F^C
- \frac{\partial_i\partial_j\partial_{\bar k}\Omega_0}{\Omega}
\hat F^j \hat F^{k*}
- \frac{\partial_i\partial_j \Omega_0}{\Omega}
\hat F^j \hat F^{C*} = 0,
\eea
at a non-supersymmetric minimum of scalar potential with vanishing vacuum energy
density $V_{\rm tot}=0$, where $\hat F^{i,C}$ are F-components rescaled
by the gravitino mass
\bea
\hat F^i \equiv \frac{F^i}{m^*_{3/2}},
\qquad
\hat F^C \equiv \frac{1}{m^*_{3/2}}\frac{F^C}{C_0}.
\eea
It should be noted that the moduli F-term relation (\ref{Moduli-F-term-relation})
derived from stationary condition does not depend on the detailed structure of
sequestered sector, $\Omega_{\rm seq}$ and $W_{\rm seq}$.
Since soft terms of visible fields are generated by $F^i$ and $F^C$,
the insensitivity of moduli F-term relation to the sequestered sector physics
is crucial for identifying the dominant source of soft terms in the visible sector.
The relation (\ref{Moduli-F-term-relation}) would allow us to determine, without
a detailed information about the sequestered sector, the relative importance of
moduli mediation compared to the anomaly mediation which is always present
in supergravity.
In the next section, by using this relation between moduli F-term VEVs, we will
examine the pattern of soft terms in generic scenario where the moduli are
stabilized by adding non-perturbative effects.

For the sequestered uplifting scenario, uplifting procedure can be also described
within an effective lagrangian that is obtained by integrating out the sequestered
sector.
Then, as constructed from the underlying theory (\ref{sequestering}), the effective
theory should reproduce the moduli F-term relation (\ref{Moduli-F-term-relation}) which
is insensitive to the sequestered sector physics.
The effects of sequestered sector fields are encoded in constant operators
appearing in the resultant effective lagrangian
\bea
{\cal L}_{\rm eff} &=&
\int d^2\theta d^2\bar\theta C\bar C
\left( -3\Omega_0(T^i,\bar T^i) + P_0
+ \bar \theta^2 \frac{C^2}{\bar C} P_1
+ \theta^2 \frac{\bar C^2}{C} P^*_1
+ \theta^2 \bar\theta^2 C\bar C P_2 \right)
\nonumber \\
&&
+\, \left\{ \int d^2\theta C^3
\left( W_0(T^i) + H_0 + \theta^2 \frac{\bar C^2}{C} H_1 \right)
+ {\rm h.c.} \right\},
\eea
where spurion operators $P_{1,2}$ and $H_1$ represent the spontaneous SUSY breaking
effects in sequestered sector, and their $C$-dependence is determined from the
fact that the combination $\theta^2 \bar C^2/C$ is invariant under the global
super-Weyl transformations.
The operators $P_{0,1,2}$ and $H_{0,1}$ are $T^i$-independent constants
because they are originated from the sequestered sector physics which is completely
decoupled in the absence of gravitational effects.
Hence, by appropriate redefinition of K\"ahler and superpotential, the effective
lagrangian is written as
\bea
\label{effective-action}
{\cal L}_{\rm eff} &=&
\int d^2\theta d^2\bar\theta
C\bar C \left( -3\Omega(T^i,\bar T^i)
- \theta^2\bar\theta^2 C\bar C {\cal P}_{\rm lift} \right)
+ \left\{ \int d^2\theta C^3 W(T^i)
+ {\rm h.c.} \right\},
\eea
in which the simple constant shifts of K\"ahler and superpotential have been
taken into account
\bea
\Omega(T^i,\bar T^i) = \Omega_0(T^i,\bar T^i) - \frac{1}{3}P_0
\quad{\rm and}\quad
W(T^i) = W_0(T^i) + H_0 + P_1,
\eea
whereas the SUSY breaking effects by $Z^\alpha$ and $\tilde V^a$ are
encoded in a D-type spurion operator
\bea
{\cal P}_{\rm lift} &=& -P_2 - H_1 - H^*_1,
\eea
which appears basically when the F-term of sequestered matter fields develops
a VEV, as the D-terms vanish if $D_\alpha W=0$ for all $Z^\alpha$.
We note that, in the effective action, this spurion operator ${\cal P}_{\rm lift}$
mimics explicit SUSY breaking in a sequestered sector and provides a KKLT-type
uplifting potential\footnote{
In the KKLT flux compactification \cite{Kachru:2003aw}, an anti-brane is
stabilized at the end of a warped throat, while the visible brane is supposed
to be located at a region where the warping is negligible.
The sequestering is then achieved for the spatially separated two
branes through the warping, while uplifting potential is provided by explicit
SUSY breaking on the anti-brane.}
given by
\bea
\label{uplifting-potential}
V_{\rm lift} &=& e^{2K/3}{\cal P}_{\rm lift},
\eea
in the Einstein frame.
For sequestered uplifting models with spontaneously broken SUSY, the moduli
stabilization is therefore expected to have qualitatively same features as those
in the KKLT scenario \cite{Kachru:2003aw,Choi:2004sx,Choi:2005ge}.
Including the uplifting potential (\ref{uplifting-potential}), the vacuum configuration
is now determined by solving $\partial_i V_{\rm tot}=\partial_i(V_F+V_{\rm lift})=0$
where
\bea
\partial_i V_{\rm lift} &=& -2\frac{\partial_i \Omega}{\Omega}V_{\rm lift},
\eea
from which it is straightforward to obtain the same result for moduli F-term VEVs
as (\ref{Moduli-F-term-relation}) for vanishing vacuum energy,
$V_{\rm tot}=V_F+V_{\rm lift}=0$.
The moduli F-term relation remains unmodified in the effective theory, as deduced
from the fact that it is the consequence of full theory (\ref{sequestering}) and
insensitive to the sequestered sector physics.
Therefore, the SUSY breaking effects producing visible soft terms can be studied
consistently by using the relation (\ref{Moduli-F-term-relation}) in the effective
theory (\ref{effective-action}), where the spontaneous SUSY breaking effects from
sequestered sector are represented by a KKLT-type uplifting operator
${\cal P}_{\rm lift}$.

\section{Moduli stabilization}
Non-perturbative effects, such as gaugino condensations \cite{Derendinger:1985kk},
are considered as natural sources of moduli stabilization.
In this section, including a sequestered sector, we discuss the pattern of soft
terms in generic scenario where the moduli $T^i$ are fixed by adding
non-perturbative corrections to superpotential
\bea
\label{Wnp}
W_0(T^i) &=& \sum_i A_i e^{-a_i T^i},
\eea
where $A_i={\cal O}(1)$ in the unit with $M_{Pl}=1$, and moduli are defined through
the exponent.
Provided that the K\"ahler potential for moduli $T^i$ and its derivatives do not
introduce hierarchically large numbers, the vanishing vacuum energy forces moduli
F-terms to be at most of order of gravitino mass
\bea
\label{const1}
\left| \frac{1}{m^*_{3/2}}\frac{F^i}{T^i} \right| &\leq& {\cal O}(1).
\eea
Furthermore, in order for supersymmetry to resolve the hierarchy problem,
soft parameters have to be of TeV scale \cite{Nilles:1983ge}.
Since soft terms of visible fields are induced by moduli and anomaly mediations,
the low energy SUSY requires F-term VEVs to satisfy
\bea
\label{const2}
\left|\frac{1}{4\pi^2}\frac{F^C}{C_0}\right|
+ \sum_i \left|\frac{F^i}{T^i}\right| &=& {\cal O}(1)\,{\rm TeV},
\eea
where we have used the fact that anomaly mediated contributions are suppressed
by a loop factor.
As long as the non-perturbative superpotential is the main source of moduli
fixing, the associated moduli would be fixed according to
$A_i e^{-a_i T^i} \sim W$.
This implies $\langle a_i T^i \rangle \approx \ln(M_{Pl}/m_{3/2})\gg1$,
because the moduli appear non-perturbatively in the superpotential.
Hence, combined with the constraints on F-terms (\ref{const1}) and (\ref{const2}),
the moduli F-term relation (\ref{Moduli-F-term-relation}) gives
\bea
\label{NP-F-relation}
\frac{F^i}{T^i} \simeq \frac{2}{a_iT^i}\frac{F^C}{C_0}
\approx \frac{2}{\ln(M_{Pl}/m_{3/2})}\frac{F^C}{C_0}
\qquad{\rm with}\qquad m_{3/2}\geq {\cal O}(1)\,{\rm TeV}
\eea
at a non-supersymmetric minimum with vanishing vacuum energy.
We stress that the above result does not depend on the detailed form of
sequestered sector, $\Omega_{\rm seq}$ and $W_{\rm seq}$.
If there are no SUSY breaking fields other than $T^i$, the moduli F-term VEVs
are found to be $F^i/T^i={\cal O}(m_{3/2}/\ln(M_{Pl}/m_{3/2}))$ from
(\ref{FD-terms}) and (\ref{NP-F-relation}), and thus are too small to cancel
the negative vacuum energy of ${\cal O}(m^2_{3/2}M^2_{Pl})$ in supergravity
potential.
This indicates that the uplifting effects from sequestered sector are responsible
for the construction of a dS vacuum with nearly vanishing cosmological constant.
It should be also noted that, owing to the suppression of moduli F-terms compared
to $F^C$ (\ref{NP-F-relation}), the moduli mediation would give comparable
contributions to soft terms as the loop-induced anomaly mediation.
Consequently, the soft terms of visible fields are predicted to take the pattern
of mirage mediation \cite{Choi:2004sx,Choi:2005ge,Choi:2005uz,Endo:2005uy}
at low energies.
In the sequestered uplifting scenario with spontaneously broken SUSY, mirage mediation
is a natural consequence of non-perturbative moduli stabilization where the moduli
appear non-perturbatively in superpotential.
Moreover, the source of uplifting effects can be regarded as KKLT-like explicit SUSY
breaking when the sequestered sector is integrated out.
These features can be understood essentially from the fact that the moduli F-term
relation (\ref{Moduli-F-term-relation}) is insensitive to the sequestered sector
physics.

To see the phenomenological aspects, we consider a simple case with a single
K\"ahler modulus $T$.
In the presence of a sequestered sector, the effective supergravity action
is described by
\bea
K = -3\ln\left(T+\bar T
+ \Omega_{\rm seq}(Z^\alpha,\bar Z^\alpha,\tilde V^a)\right),
\qquad
W = Ae^{-aT} + W_{\rm seq}(Z^\alpha),
\eea
where the non-perturbative superpotential is added to stabilize $T$,
and $A={\cal O}(1)$.
Applying the relation (\ref{Moduli-F-term-relation}) to this model, the VEV of modulus
F-term is found to be exactly given by
\bea
\label{model-F-relation}
\frac{F^T}{T} &=& -\frac{2}{T}
\frac{\partial_T W}{\partial^2_T W}\frac{F^C}{C_0}
= \frac{2}{a T}\frac{F^C}{C_0},
\eea
which is totally independent of the detailed structure of sequestered sector.
Here the sequestered uplifting potential is adjusted to get a vanishing vacuum
energy density.
Due to the suppression of modulus F-term, the low energy values of soft parameters are
expected to take the pattern of mirage mediation, where the mirage messenger scale
\cite{Choi:2005uz} is determined by the ratio between modulus and anomaly mediations.
The mirage messenger scale does not correspond to a physical threshold scale.
Its appearance reflects the fact that the anomaly mediated contribution cancels
precisely the renormalization-group evolved part of soft parameters.
Since the non-perturbative superpotential is supposed to be responsible for the
stabilization of $T$, the modulus would have a VEV as
$\langle aT \rangle\approx \ln(M_{Pl}/m_{3/2})$.
Using this property for non-perturbative moduli stabilization, the mirage messenger scale
is then estimated easily from (\ref{model-F-relation}), irrespectively of the detailed
form of sequestered uplifting.
Mirage mediation models can naturally avoid the SUSY CP and flavor problems
as a consequence of approximate scaling and axionic shift symmetries
\cite{Choi:2006za,Choi:1993yd,Conlon:2006tj}.
It has been also noticed that the mirage mediation has a distinctive pattern of low
energy soft parameters and interesting phenomenological implications
\cite{Choi:2005uz,Endo:2005uy}.

\section{Conclusion}
We have studied the pattern of soft terms in the generic scenario of moduli
stabilization, where a dS vacuum is constructed by combining a sequestered sector.
Fields in the sequestered sector give contributions to spontaneous SUSY
breaking, which achieve uplifting effects.
The cosmological constant can be then adjusted to be arbitrarily small.
In this framework, the addition of uplifting sector causes no flavor
violations because the sequestering forbids cross-couplings between the visible
and uplifting fields.

In supergravity, the moduli configuration minimizing scalar potential is
generally affected by the sequestered sector through gravitational effects.
Nonetheless, written in terms of SUSY breaking quantities, stationary conditions
are found to have the decoupled structure.
As a result, we obtain the relation between moduli F-term VEVs which does not
depend on the detailed form of sequestered uplifting.
This relation is crucial for determining the relative importance of moduli
mediation compared to anomaly mediation which is always present
in supergravity.
For the study of phenomenological aspects, it is important to know the dominant
source of soft terms of visible fields.

It is natural to add non-perturbative effects as a source of moduli fixing.
Then, from the moduli F-term relation, it is found that moduli stabilization by
non-perturbative superpotential naturally leads to mirage mediation,
while a dS vacuum is constructed via the sequestered uplifting.
The qualitative features are same as those in the KKLT scenario.
These results can be understood from the fact that the moduli F-term relation is
insensitive to the sequestered sector physics, and a KKLT-type uplifting
potential arises in the effective theory when the sequestered sector is
integrated out.

\vspace{5mm}
\noindent{\large\bf Acknowledgments}
\vspace{5mm}

The author would like to thank K. Choi for helpful discussions.


\begin{thebibliography}{99}
\bibitem{Kachru:2003aw}
  S.~Kachru, R.~Kallosh, A.~Linde and S.~P.~Trivedi,
  Phys.\ Rev.\  D {\bf 68}, 046005 (2003)
  [arXiv:hep-th/0301240].

\bibitem{Lebedev:2006qq}
  O.~Lebedev, H.~P.~Nilles and M.~Ratz,
  Phys.\ Lett.\  B {\bf 636}, 126 (2006)
  [arXiv:hep-th/0603047];
  E.~Dudas, C.~Papineau and S.~Pokorski,
  JHEP {\bf 0702}, 028 (2007)
  [arXiv:hep-th/0610297];
  H.~Abe, T.~Higaki, T.~Kobayashi and Y.~Omura,
  Phys.\ Rev.\  D {\bf 75}, 025019 (2007)
  [arXiv:hep-th/0611024];
  R.~Kallosh and A.~Linde,
  JHEP {\bf 0702}, 002 (2007)
  [arXiv:hep-th/0611183];
  O.~Lebedev, V.~Lowen, Y.~Mambrini, H.~P.~Nilles and M.~Ratz,
  JHEP {\bf 0702}, 063 (2007)
  [arXiv:hep-ph/0612035];
  M.~Gomez-Reino and C.~A.~Scrucca,
  JHEP {\bf 0708}, 091 (2007)
  [arXiv:0706.2785 [hep-th]];
  H.~Abe, T.~Higaki and T.~Kobayashi,
  Phys.\ Rev.\  D {\bf 76}, 105003 (2007)
  [arXiv:0707.2671 [hep-th]];
  P.~Brax, A.~C.~Davis, S.~C.~Davis, R.~Jeannerot and M.~Postma,
  JHEP {\bf 0709}, 125 (2007)
  [arXiv:0707.4583 [hep-th]].

\bibitem{Burgess:2003ic}
  C.~P.~Burgess, R.~Kallosh and F.~Quevedo,
  JHEP {\bf 0310}, 056 (2003)
  [arXiv:hep-th/0309187];
  G.~Villadoro and F.~Zwirner,
  Phys.\ Rev.\ Lett.\  {\bf 95}, 231602 (2005)
  [arXiv:hep-th/0508167];
  A.~Achucarro, B.~de Carlos, J.~A.~Casas and L.~Doplicher,
  JHEP {\bf 0606}, 014 (2006)
  [arXiv:hep-th/0601190];
  S.~L.~Parameswaran and A.~Westphal,
  JHEP {\bf 0610}, 079 (2006)
  [arXiv:hep-th/0602253];
  E.~Dudas and Y.~Mambrini,
  JHEP {\bf 0610}, 044 (2006)
  [arXiv:hep-th/0607077].

\bibitem{Choi:2006bh}
  K.~Choi and K.~S.~Jeong,
  JHEP {\bf 0608}, 007 (2006)
  [arXiv:hep-th/0605108].

\bibitem{Randall:1998uk}
  L.~Randall and R.~Sundrum,
  Nucl.\ Phys.\  B {\bf 557}, 79 (1999)
  [arXiv:hep-th/9810155].

\bibitem{Luty:2000ec}
  M.~A.~Luty and R.~Sundrum,
  Phys.\ Rev.\  D {\bf 64}, 065012 (2001)
  [arXiv:hep-th/0012158];
  S.~Kachru, L.~McAllister and R.~Sundrum,
  arXiv:hep-th/0703105.

\bibitem{Schmaltz:2006qs}
  M.~Schmaltz and R.~Sundrum,
  JHEP {\bf 0611}, 011 (2006)
  [arXiv:hep-th/0608051].

\bibitem{Maldacena:1997re}
  J.~M.~Maldacena,
  Adv.\ Theor.\ Math.\ Phys.\  {\bf 2}, 231 (1998)
  [Int.\ J.\ Theor.\ Phys.\  {\bf 38}, 1113 (1999)]
  [arXiv:hep-th/9711200].

\bibitem{Choi:2004sx}
  K.~Choi, A.~Falkowski, H.~P.~Nilles, M.~Olechowski and S.~Pokorski,
  JHEP {\bf 0411}, 076 (2004)
  [arXiv:hep-th/0411066].

\bibitem{Choi:2005ge}
  K.~Choi, A.~Falkowski, H.~P.~Nilles and M.~Olechowski,
  Nucl.\ Phys.\  B {\bf 718}, 113 (2005)
  [arXiv:hep-th/0503216].

\bibitem{Choi:2005uz}
  K.~Choi, K.~S.~Jeong and K.~i.~Okumura,
  JHEP {\bf 0509}, 039 (2005)
  [arXiv:hep-ph/0504037].

\bibitem{Endo:2005uy}
  M.~Endo, M.~Yamaguchi and K.~Yoshioka,
  Phys.\ Rev.\  D {\bf 72}, 015004 (2005)
  [arXiv:hep-ph/0504036];
  A.~Falkowski, O.~Lebedev and Y.~Mambrini,
  JHEP {\bf 0511}, 034 (2005)
  [arXiv:hep-ph/0507110].

\bibitem{Cremmer:1982en}
  E.~Cremmer, S.~Ferrara, L.~Girardello and A.~Van Proeyen,
  Nucl.\ Phys.\  B {\bf 212}, 413 (1983).

\bibitem{Pomarol:1999ie}
  A.~Pomarol and R.~Rattazzi,
  JHEP {\bf 9905}, 013 (1999)
  [arXiv:hep-ph/9903448].

\bibitem{Giudice:1998xp}
  G.~F.~Giudice, M.~A.~Luty, H.~Murayama and R.~Rattazzi,
  JHEP {\bf 9812}, 027 (1998)
  [arXiv:hep-ph/9810442].

\bibitem{Derendinger:1985kk}
  J.~P.~Derendinger, L.~E.~Ibanez and H.~P.~Nilles,
  Phys.\ Lett.\  B {\bf 155}, 65 (1985);
  M.~Dine, R.~Rohm, N.~Seiberg and E.~Witten,
  Phys.\ Lett.\  B {\bf 156}, 55 (1985);
  H.~P.~Nilles,
  arXiv:hep-th/0402022.

\bibitem{Nilles:1983ge}
  H.~P.~Nilles,
  Phys.\ Rept.\  {\bf 110}, 1 (1984);
  H.~E.~Haber and G.~L.~Kane,
  Phys.\ Rept.\  {\bf 117}, 75 (1985).

\bibitem{Choi:2006za}
  K.~Choi and K.~S.~Jeong,
  JHEP {\bf 0701}, 103 (2007)
  [arXiv:hep-th/0611279].

\bibitem{Choi:1993yd}
  K.~Choi,
  Phys.\ Rev.\ Lett.\  {\bf 72}, 1592 (1994)
  [arXiv:hep-ph/9311352].

\bibitem{Conlon:2006tj}
  J.~P.~Conlon, D.~Cremades and F.~Quevedo,
  JHEP {\bf 0701}, 022 (2007)
  [arXiv:hep-th/0609180];
  J.~P.~Conlon, S.~S.~Abdussalam, F.~Quevedo and K.~Suruliz,
  JHEP {\bf 0701}, 032 (2007)
  [arXiv:hep-th/0610129].




\end{thebibliography}
\end{document}